# Optimal Order of Decoding for Max-Min Fairness in $K$-User Memoryless Interference Channels

Mohammad Ali Maddah-Ali, Hajar Mahdavi-Doost, and Amir K. Khandani

Coding & Signal Transmission Laboratory(www.cst.uwaterloo.ca)

Dept. of Elec. and Comp. Eng., University of Waterloo

Waterloo, ON, Canada, N2L 3G1

e-mail: {mohammad, hajar, khandani}@cst.uwaterloo.ca

**Abstract**

A $K$-user memoryless interference channel is considered where each receiver sequentially decodes the data of a subset of transmitters before it decodes the data of the designated transmitter. Therefore, the data rate of each transmitter depends on (i) the subset of receivers which decode the data of that transmitter, (ii) the decoding order, employed at each of these receivers. In this paper, a greedy algorithm is developed to find the users which are decoded at each receiver and the corresponding decoding order such that the minimum rate of the users is maximized. It is proven that the proposed algorithm is optimal.

## I. INTRODUCTION

Wireless technology has been advancing at an exponential rate, due to increasing expectations for multi-media services. This, in turn, necessitates the development of novel techniques of signaling with high spectral efficiency. Channel sharing is known as an effective scheme to increase the spectral efficiency and coverage in the wireless systems. The main source of impairment in such systems is the interference among the links. These systems are known with the general name of interference channels.

This work is financially supported by Communications and Information Technology Ontario (CITO), Nortel Networks, and National Sciences and Engineering Research Council of Canada (NSERC).

The interference channel was first introduced by Shannon [1]. In [2], it is shown that in the Gaussian interference channels, very strong interference amounts to no interference at all. In [3]–[5], the result of [2] is extended to general discrete interference channels with strong interference. In [6], [7], the capacity of degraded interference channels is investigated. The best result on the capacity region of the interference channels is introduced in [5]. In the scheme presented in [5], each transmitter splits its message into two independent massages, one is private which is only decodable by the intended receiver and the other is common which is decodable at both receivers.

A lot of research efforts have been devoted to the problem of fairness in the interference channels. In [8], $K$-user Gaussian interference channels without any constraint on the transmit powers are considered and the maximum signal-to-interference-plus-noise-ratio (SINR) that all the transmitters can attain simultaneously is computed. The result in [8] is formulated as the inverse of the Perron-Frobenius eigenvalue (see [9]) of a non-negative matrix. Recently in [10], the result of [8] is generalized to the case where the power of the transmitters are constrained. In [11], the problem of spectrum sharing in unlicensed bands is investigated. It is shown that in a $K$-user interference channel, any rate vector inside the rate region is achievable with a piece-wise constant power allocation over $2K$ bandwidth intervals. In addition, it is investigated whether fairness and efficiency can be attained if the users follow a selfish spectrum sharing strategy. Generally in the literature, including [8], [10], [11], it is assumed that each receiver only decodes the data of the designated transmitter, while the signals coming from other transmitters are treated as interference.

In this paper, we consider a $K$-user memoryless interference channel, where each receiver sequentially decodes the data of a subset of transmitters before it decodes the data of the designated transmitter. Since part of the interference is canceled out, this system can potentially achieve higher data rate. In this system, the data rate of each transmitter depends on (i) the subset of receivers which decode the data of that transmitter, (ii) the decoding order employed at each receiver which decodes the data of that transmitter. The main objective of this paper is to find the set of transmitters which are decoded at each receiver and the corresponding order of decoding such that the minimum rate of the users is maximized. A simple greedy algorithm is proposed and proven to be optimal. We established similar result for the memoryless multi-access channels in [12].

## II. Problem Formulation

We focus on a $K$-user memoryless interference channel modeled by

$$\Pr(y_1, y_2, \ldots, y_K | x_1, x_2, \ldots, x_K). \tag{1}$$

It is assumed that user $k$, $k \in E = \{1, 2, \ldots, K\}$, utilizes the codebook $\mathcal{C}^{(k)}$, with the input distribution $\Pr(x_k)$. Receivers have the possibility of successive decoding. Each receiver decodes the data of some of the users in a specific order and then it decodes the data of the designated transmitter. For the sake of brevity, we say "user $k$ is decoded at receiver $j$", instead of saying "the data of the user $k$ is decoded at receiver $j$".

The order of decoding at receiver $j$ is denoted by the permutation $\boldsymbol{\pi}^{(j)} = (\pi^{(j)}(1), \pi^{(j)}(2), \ldots, \pi^{(j)}(K))$ of the set $E$. Receiver $j$ first decodes user $\pi^{(j)}(K)$, then user $\pi^{(j)}(K-1)$, and so forth until it decodes the data of the designated transmitter (See Fig. 1). In the permutation $\boldsymbol{\pi}^{(j)}$, if $l > i$ ($l < i$), we say user $\pi^{(j)}(l)$ is located before (after) user $\pi^{(j)}(l)$, which means that at receiver $j$, user $\pi^{(j)}(l)$ is decoded before (after) user $\pi^{(j)}(i)$. Apparently, the users located after user $j$ in the permutation $\boldsymbol{\pi}^{(j)}$ are not decoded at receiver $j$. The orders of decoding at all receivers, i.e., $\boldsymbol{\pi}^{(1)}, \boldsymbol{\pi}^{(2)}, \ldots, \boldsymbol{\pi}^{(K)}$, are denoted by $\Pi$.

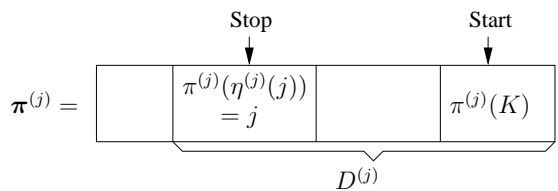

Fig. 1. Order of Decoding at Receiver $j$

**Definition 1** *The vector $\boldsymbol{\eta}^{(k)}$ is defined such that $\eta^{(k)}(j)$ shows the position of user $k$ in $\boldsymbol{\pi}^{(j)}$, therefore,*

$$\pi^{(j)}(\eta^{(k)}(j)) = k.$$

**Definition 2** *The set $D^{(j)}$ is defined as the set of users which are decoded at receiver $j$, i.e.,*

$$D^{(j)} = \{\pi^{(j)}(\eta^{(j)}(j)), \pi^{(j)}(\eta^{(j)}(j)+1), \ldots, \pi^{(j)}(K)\}. \tag{2}$$

Note that $\pi^{(j)}(\eta^{(j)}(j))$ is equal to $j$, which is the last user, decoded at receiver $j$. The users located after user $j$ in $\boldsymbol{\pi}^{(j)}$ are not decoded at receiver $j$.

**Definition 3** *The set $E^{(k)}$ is defined as the set of receivers which decode user $k$. Apparently, $k \in E^{(k)}$.*

Receiver $j$ and the transmitter in $D^{(j)}$ can be considered as a multi-access channel, while the contributions of the users in $E - D^{(j)}$ are treated as interference. Regarding the order of decoding applied at receiver $j$, the rate of user $k$, $k \in D^{(j)}$, is upper-bounded by

$$r_k \leq \tag{3}$$
$$I(y_j; x_k | x_{\pi^{(j)}(\eta^{(k)}(j)+1)}, x_{\pi^{(j)}(\eta^{(k)}(j)+2)}, \ldots, x_{\pi^{(j)}(K)}).$$

Note that $\{x_{\pi^{(j)}(\eta^{(k)}(j)+1)}, x_{\pi^{(j)}(\eta^{(k)}(j)+2)}, \ldots, x_{\pi^{(j)}(K)}\}$ is the set of users decoded before user $k$ at receiver $j$.

Therefore, if the decoding orders $\Pi$ are employed at the receivers, the maximum possible value for $r_k$, denoted by $r_k(\Pi)$, is obtained by,

$$r_k(\Pi) = \min_{j,\ j \in E^{(k)}} \tag{4}$$
$$I(y_j; x_k | x_{\pi^{(j)}(\eta^{(k)}(j)+1)}, x_{\pi^{(j)}(\eta^{(k)}(j)+2)}, \ldots, x_{\pi^{(j)}(K)}). \tag{5}$$

The objective of this paper is to find the optimal decoding orders $\pi^{(k)}$, $k = 1, \ldots, K$, such that the minimum of $r_k(\Pi)$, $k = 1, \ldots, K$, is maximized.

Note that there are $\left(\sum_{i=1}^{K} \frac{K!}{i!}\right)^K$ possible choices for the decoding orders, and it is prohibitively complex to find the optimal answer through the exhaustive search.

## III. Preliminaries

**Definition 4** *[13, Ch. 18] Let $E = \{1, 2, \ldots, K\}$ and $f : 2^E \longrightarrow \mathcal{R}_+$ be a set function. $f$ is called a rank function, if it satisfies the following conditions,*

$$(normalized) \quad f(\emptyset) = 0, \tag{6}$$
$$(increasing) \quad f(S) \leq f(T) \text{ if } S \subset T, \tag{7}$$
$$(submodular) \quad f(S) + f(T) \geq f(S \cap T) + f(S \cup T). \tag{8}$$

We define the set function $f^{(j)}$ as

$$f^{(j)}(S) = I\left(y_j; \{x_i, i \in S\} | \{x_i, i \in S^c\}\right), \ \forall S \subset E. \tag{9}$$

It is proven that $f^{(j)}(S)$ is a rank function [14]. In addition, it is easy to see that (3) and (4) are respectively rewritten as

$$r_k \leq f^{(j)}(x_{\pi^{(j)}(1)}, x_{\pi^{(j)}(2)}, \ldots, x_{\pi^{(j)}(\eta^{(k)}(j))}) \tag{10}$$
$$- f^{(j)}(x_{\pi^{(j)}(1)}, x_{\pi^{(j)}(2)}, \ldots, x_{\pi^{(j)}(\eta^{(k)}(j)-1)}).$$

and

$$r_k(\Pi) = \min_{j,\ j \in E^{(k)}} \tag{11}$$
$$f^{(j)}(x_{\pi^{(j)}(1)}, x_{\pi^{(j)}(2)}, \ldots, x_{\pi^{(j)}(\eta^{(k)}(j))})$$
$$- f^{(j)}(x_{\pi^{(j)}(1)}, x_{\pi^{(j)}(2)}, \ldots, x_{\pi^{(j)}(\eta^{(k)}(j)-1)}).$$

## IV. Algorithm

In this section, we develop an algorithm to specify the optimal decoding orders. In the proposed algorithm, the decoding order for each receiver is determined in a greedy fashion, independent of the decoding orders selected for the other receivers. While this algorithm has a very low complexity, we prove that the resulting decoding orders are optimal.

**Algorithm I**

For each receiver $j$, $j \in E$,

1) Set $\alpha = K$, $D^{*(j)} = \varnothing$.
2) Set $\pi^{*(j)}(\alpha)$ as
$$\pi^{*(j)}(\alpha) = \arg\min_{z \in E, z \notin \mathcal{S}} f^{(j)}\left(E - D^{*(j)} - \{z\}\right). \tag{12}$$
3) Set $D^{*(j)} \longleftarrow D^{*(j)} \cup \{\pi^{*(j)}(\alpha)\}$ and $\alpha \longleftarrow \alpha - 1$. If $\alpha \geq 1$ and $\pi^{*(j)}(\alpha+1) \neq k$, then go to step two, otherwise go to the next step.
4) If $\alpha \neq 0$, randomly allocate the the entries of $E - D^{*(j)}$ to $\pi^{(j)}(1), \pi^{(j)}(2), \ldots, \pi^{(j)}(\alpha)$.

The following theorem proves the optimality of the proposed algorithm.

**Theorem 1** *Let $\big(r_1(\Pi^*), r_2(\Pi^*), \ldots, r_K(\Pi^*)\big)$ be the rate vector corresponding to the decoding orders $\pi^{*(1)}, \pi^{*(2)}, \ldots, \pi^{*(K)}$. Then for the rate vector $\big(r_1(\Pi), r_2(\Pi), \ldots, r_K(\Pi)\big)$ corresponding to any decoding orders $\pi^{(1)}, \pi^{(2)}, \ldots, \pi^{(K)}$, we have*

$$\min_i r_i(\Pi^*) \geq \min_i r_i(\Pi). \tag{13}$$

*Proof:* Let $\boldsymbol{\eta}^{*(j)}$ and $E^{(*j)}$ respectively be $\boldsymbol{\eta}^{(j)}$ and $E^{(j)}$ corresponding to the decoding orders obtained by the algorithm. Assume that user $\theta$ has the minimum rate among the users, where the decoding orders $\boldsymbol{\pi}^{*(1)}, \boldsymbol{\pi}^{*(2)}, \ldots, \boldsymbol{\pi}^{*(K)}$ are employed at the receivers. Therefore, regarding (11), $\exists j \in E^{(*\theta)}$ such that

$$r_\theta(\Pi^*) = f^{(j)}\big(x_{\pi^{*(j)}(1)}, x_{\pi^{*(j)}(2)}, \ldots, x_{\pi^{(*j)}(\eta^{(*\theta)}(j))}\big) - f^{(j)}\big(x_{\pi^{(*j)}(1)}, x_{\pi^{(*j)}(2)}, \ldots, x_{\pi^{(*j)}(\eta^{(*\theta)}(j)-1)}\big) \tag{14}$$

In other words, among the receives which decode user $\theta$, the receiver $j$ imposes the dominant upper-bound on the data rate of the user $\theta$. For now, we assume that $\theta \neq j$. Similar arguments are used to prove the optimality of the algorithm for the case that $\theta = j$.

In what follows, we prove that if the decoding orders $\boldsymbol{\pi}^{*(1)}, \boldsymbol{\pi}^{*(2)}, \ldots, \boldsymbol{\pi}^{*(K)}$ are permuted to generate new decoding orders, then the minimum rate of users is not greater than $r_\theta(\Pi^*)$.

**Case 1.** *Choosing arbitrary permutations for $\boldsymbol{\pi}^{(l)}$, $l \in E, l \neq j$:* Assume that arbitrary decoding orders are chosen for the receivers $l$, $l \in E$ and $l \neq j$, while the user $j$ is employed $\boldsymbol{\pi}^{*(j)}$ as the decoding order. Then user $\theta$ is still decoded at receiver $j$, in the order determined by $\pi^{*(j)}$. Therefore, according to (11), the rate of user $\theta$ is still upper-bounded by the right-hand side of (14), which is $r_\theta(\Pi^*)$. Consequently, if the new decoding orders are employed, the minimum rate of the users is less than or equal to $r_\theta(\Pi^*)$.

Before starting the other cases, we define two sets:

- The set of users located after user $\theta$ in the permeation $\pi^{*(j)}$,

$$\Phi^{*(j)} = \{\pi^{*(j)}(1), \ldots, \pi^{*(j)}(\eta^{(*\theta)}(j) - 1)\}. \tag{15}$$

Note that $j \in \Phi^{*(j)}$. In addition, some of the users in $\Phi^{*(j)}$ are not decoded at receiver $j$.

- The set of users decoded before user $\theta$ at receiver $j$ according to the permutation $\pi^{*(j)}$:

$$\Psi^{*(j)} = \{\pi^{*(j)}(\eta^{(*\theta)(j)} + 1), \ldots, \pi^{*(j)}(K)\} \tag{16}$$

Therefore, according to (14), we have

$$r_\theta(\Pi^*) = f^{(j)}(\Phi^{*(j)} \cup \{\theta\}) - f^{(j)}(\Phi^{*(j)}). \tag{17}$$

**Case 2.** *Permutation in $\Phi^{*(j)}$ and $\Psi^{*(j)}$, choosing arbitrary permutations for $\boldsymbol{\pi}^{(l)}$, $l \in E, l \neq j$ (see Fig. 2):* Assume that the order of users in $\Phi^{*(j)}$ and $\Psi^{*(j)}$ are permuted to generate a new decoding order $\boldsymbol{\pi}^{(j)}$ for receiver $j$. Note that in the new permutation $\boldsymbol{\pi}^{(j)}$, the set of users located after and before user $\theta$ are still $\Phi^{*(j)}$ and $\Psi^{*(j)}$. Also assume that for the rest of receivers, arbitrary decoding orders are chosen. In this case, in $\boldsymbol{\pi}^{(j)}$, user $j$ is still located after user $\theta$ and therefore, user $\theta$ is decoded at receiver $j$. In addition, according to (11), the rate of user $\theta$ is still less than $f^{(j)}(\Phi^{*(j)} \cup \{\theta\}) - f^{(j)}(\Phi^{*(j)})$, to be decodable at receiver $j$. Therefore, if the new decoding orders are employed, the minimum rate of the users is less that or equal to $r_\theta(\Pi^*)$.

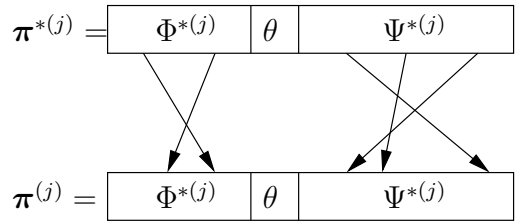

Fig. 2. Case 2. Permutation in $\Phi^{*(j)}$ and $\Psi^{*(j)}$.

**Case 3.** *Moving a subset of users from $\Psi^{*(j)}$ to $\Phi^{*(j)}$, choosing arbitrary permutations for $\boldsymbol{\pi}^{(l)}$, $l \in E, l \neq j$ (See Fig 3):* Assume a set $\Upsilon$ of users, $\Upsilon \subset \Psi^{*(j)}$, is moved from $\Psi^{*(j)}$ to $\Phi^{*(j)}$ to generate a new decoding order $\boldsymbol{\pi}^{(j)}$ for receiver $j$. Note that in the permutation $\boldsymbol{\pi}^{(j)}$, the position of user $\theta$ is still before user $j$, which means that user $\theta$ is decoded at receiver $j$. Assume that arbitrary permutations are chosen for the other receivers. According to (11), if the new decoding orders are employed, the rate of user $\theta$ is less than or equal to,

$$r_\theta(\Pi) \leq f(\Phi^{*(j)} \cup \Upsilon \cup \{\theta\}) - f(\Phi^{*(j)} \cup \Upsilon), \tag{18}$$

to be decodable at receiver $j$, regardless of the decoding orders chosen for the other receivers.

Using (8), we have,

$$f(\Phi^{*(j)} \cup \{\theta\}) + f(\Phi^{*(j)} \cup \Upsilon) \geq$$
$$f(\Phi^{*(j)} \cup \Upsilon \cup \{\theta\}) + f(\Phi^{*(j)}). \tag{19}$$

Using (17), (18), and (19), we conclude that $r_\theta(\Pi) \leq r_\theta(\Pi^*)$, and therefore, the minimum rate of the users in the new decoding orders is less than or equal to $r_\theta(\Pi^*)$.

Note that permuting the users located before (or after) user $\theta$ in $\boldsymbol{\pi}^{(j)}$ does not increase the rate of user $\theta$.

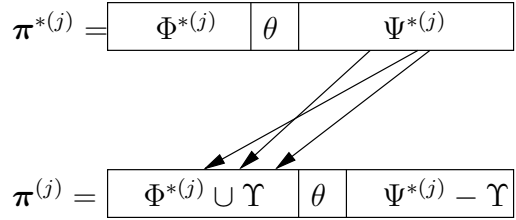

Fig. 3. Case 3. Moving a set of users from $\Psi^{*(j)}$ to the set $\Phi^{*(j)}$.

**Case 4.** *Moving one or more users from the set $\Phi^{*(j)}$ to the set $\Psi^{*(j)}$, with or without moving some users from the set $\Psi^{*(j)}$ to the set $\Phi^{*(j)}$, choosing arbitrary permutations for $\boldsymbol{\pi}^{(l)}$, $l \in E, l \neq j$(See Fig 4)*: Assume that one or more users move from $\Phi^{*(j)}$ to $\Psi^{*(j)}$ (with or without moving some users from the set $\Psi^{*(j)}$ to the set $\Phi^{*(j)}$) to generate the new permutation $\boldsymbol{\pi}^{(j)}$. As depicted in Fig. 4, assume that the user $\nu$ is positioned last in the permutation $\boldsymbol{\pi}^{(j)}$ among the users moved from $\Phi^{*(j)}$ to $\Psi^{*(j)}$ (user $\pi(1)$ is positioned first and user $\pi(K)$ is positioned last in the permutation $\boldsymbol{\pi}$). In the new permutation, user $\nu$ is located before user $j$, which means that this user is decoded at receiver $j$, otherwise, user $\nu$ is indeed user $j$ which is apparently decoded at receiver $j$.

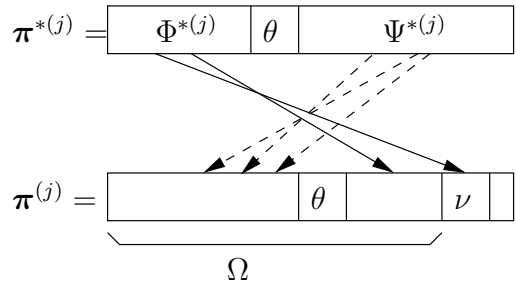

Fig. 4. Case 4. Moving one or more users from the set $\Phi^{*(j)}$ to the set $\Psi^{*(j)}$ (with or without moving some users from the set $\Psi^{*(j)}$ to the set $\Phi^{*(j)}$).

Let $\Omega$ be the set of users located after the user $\nu$ in the permutation $\boldsymbol{\pi}^{(j)}$. Using (11), and since $\nu$ is decoded at receiver $j$, the rate of user $\nu$ is upper-bounded by,

$$r_\nu(\Pi) \leq f^{(j)}(\Omega \cup \{\nu\}) - f^{(j)}(\Omega), \tag{20}$$

to be decodable at receiver $j$. It is clear that,

$$\{\theta\} \cup \Phi^{*(j)} - \{\nu\} \subset \Omega. \tag{21}$$

Using (8) with $S = \Phi^{*(j)} \cup \{\theta\}$ and $T = \Omega$, and regarding (21), we have,

$$f^{(j)}(\Omega \cup \{\nu\}) - f^{(j)}(\Omega)$$
$$\leq f^{(j)}(\Phi^{*(j)} \cup \{\theta\}) - f^{(j)}(\Phi^{*(j)} \cup \{\theta\} - \{\nu\}). \tag{22}$$

On the other hand, user $\nu$ is in the set $\Phi^{*(j)}$ in permutation $\boldsymbol{\pi}^{*(j)}$. It means that in Step 2 of the algorithm, this user has been compared with other users in the set $\Phi^{*(j)} \cup \{\theta\}$ to be located in the position $\eta^{*(\theta)}(j)$ of the permutation $\boldsymbol{\pi}^{*(j)}$, but user $\theta$ has been chosen for the position, i.e., $f^{(j)}\left(\Phi^{*(j)} \cup \{\theta\} - \{\theta\}\right) \leq f^{(j)}\left(\Phi^{*(j)} \cup \{\theta\} - \{\nu\}\right)$, therefore,

$$f^{(j)}\left(\Phi^{*(j)}\right) \leq f^{(j)}\left(\Phi^{*(j)} \cup \{\theta\} - \{\nu\}\right). \tag{23}$$

Using (17), (20), (22), and (23), we conclude that $v_\nu(\Pi) \leq v_\theta(\Pi^*)$, regardless of the decoding orders chosen for the other receivers. Therefore, if the new decoding orders are employed, the minimum rate of the users is less than or equal to $r_\theta(\Pi^*)$. Note that permuting of the users located before (or after) user $\nu$ in $\boldsymbol{\pi}^{(j)}$ does not increase the rate of user $\nu$. ∎

### A. Special Case: Gaussian Interference Channels

A Gaussian interference channel, including $K$ users, is represented by the gain matrix $\mathbf{G} = [g_{j,i}]_{K \times K}$ where $g_{j,i}$ is the power gain from transmitter $i$ to receiver $j$. A white Gaussian noise with zero mean and variance $\sigma_j^2$ is added to the received signal at receiver $j$ terminal. In this case, $f^{(j)}$, defined in (9), is written as

$$f^{(j)}(S) = \log_2\left(\sigma_j^2 + \sum_{i \in S} g_{j,i} p_i\right), \tag{24}$$

where $p_i$ denotes the power of transmitter $i$.

We can show that Algorithm I simplifies as follows. The set of users decoded at receiver $j$, $D^{*(j)}$, is equal to

$$D^{*(j)} = \{k \ : \ g_{j,k} p_k \geq g_{j,j} p_j\}. \tag{25}$$

At receiver $j$, user $i$ is decoded before user $k$ if $g_{j,i}p_i \geq g_{j,k}p_j$. Therefore, to obtain the optimal decoding order for receiver $j$, we sort $g_{j,i}p_i$, $i \in E$, decreasingly. The optimal decoding order for receiver $j$, i.e., $\boldsymbol{\pi}^{(*j)}$ is such that,

$$g_{j,\pi^{(*j)}(K)}p_{\pi^{(*j)}(K)} \geq$$

$$g_{j,\pi^{(*j)}(K-1)}p_{\pi^{(*j)}(K-1)} \geq \ldots \geq g_{j,j}p_j.$$

In addition, the set of receivers which decode user $k$, i.e., $E^{*(k)}$ is derived as,

$$E^{*(k)} = \{j \ : \ g_{j,k}p_k \geq g_{j,j}p_j\}. \tag{26}$$

In this case, the rate of user $k$ is obtained by

$$r_k(\Pi^*) = \min_{j, j \in E^{*(k)}} \log_2 \left(1 + \frac{g_{j,k}p_k}{\sigma_j^2 + \sum_{i: g_{j,k}p_k \geq g_{j,i}p_i} g_{j,i}p_i}\right). \tag{27}$$